# THE NATURE OF THE 'KNEE' IN THE COSMIC RAY ENERGY SPECTRUM

## A D Erlykin [a,b] and A W Wolfendale [b]

(a) P N Lebedev Physical Institute, Moscow, Russia
(b) Department of Physics, University of Durham, Durham, UK

### ABSTRACT

In view of recent developments attention is directed again at two aspects of the well-known 'knee' in the cosmic ray energy spectrum at 3PeV: the mass of the predominant particles at this energy and their source,. It is inevitable in a subject such as this that ideas – and conclusions – evolve. Earlier, we had used a particular acceleration model [1] and the nature of the ISM in the local ISM to infer that the particles are mainly oxygen nuclei [2]; direct measurements, when extrapolated ( by at least a decade in energy ) [3], gave a similar result. Initially, [2] no specific source was identified. More recently, however, [4] we have specified the Monogem Ring supernova remnant [5] as the likely source; this is at just the right distance and age and the energies are reasonable.

Concerning the mass composition at the knee, a quantity more difficult to determine, recent direct measurements, [6], which extend to higher energies than hitherto, show a likely flattening in the spectrum above $\sim 10^4$ GeV/nucleon for He-nuclei, a flattening which, if extrapolated to higher energies, would meet the measured spectrum in the knee region. The other nuclei do not show this feature. He-nuclei in the knee region would also be marginally more consistent with KASCADE extensive air shower data, although there are serious problems with EAS mass estimates in that experiment [7].

Concerning the acceleration, recent models (*eg* [8]) applied to the Monogem Ring SNR allow a satisfactory explanation in terms of either oxygen or helium, but with the latter being a distinct possibility and perhaps more likely.

1. **INTRODUCTION**

A variety of features, including the presence of the Galactic magnetic field and lack of direct measurement of the energy spectra of the various mass components, has led to uncertainties in cosmic ray (CR) origin, even at PeV energies (the 'knee region'). In our own model [2,4,9], the knee is largely populated by oxygen nuclei from a single recent, local source ( actually, 'oxygen' is often grouped with the less abundant, nearby, nuclei - carbon and nitrogen – i.e. as CNO ). The most comprehensive indirect study has been

Corresponding author: Erlykin A.D., Department of Physics, University of Durham, South Road, Durham DH1 3LE, UK. Phone: 44-191-334-35-80, fax: 44-191-334-58-23, e-mail: a.d.erlykin@durham.ac.uk

made by the KASCADE EAS array ([7] and see [10]) but there is ambiguity in the mass assignments due to uncertainty with the model for the high energy physics involved in the analysis. Thus, for the QGSJET interaction model a deconvolution of the results yields 'knees' as follows:

P(logE=6.5), He(logE=7.0), CNO(logE=6.5) and Fe(logE=7.1): the E-values are in GeV. With the alternative model (SIBYLL) the knees (all of which are, in fact, quite sharp), are: P(logE=6.4), He(logE=6.6), CNO(logE=6.9) and Fe(logE=7.4).

There is clearly no consensus. The component having the highest intensity in the knee region is, for QGSJET: He(logE=6.9) and for SIBYLL CNO(logE=6.9).

The knee in the spectrum is at an unusually high energy, logE = 6.75, compared with the usual 6.5. The identification of the predominant mass here is equivocal: He or CNO, although we should argue that the evidence for sharp knees in the constituent spectra – indicative of a single source – is strong; this in itself suggests an important contribution from a 'single source'.

## 2. INDIRECT EVIDENCE FAVORING PREDOMINANT He AT THE KNEE

There are several indirect indications for the predominance of helium at the knee rather than oxygen:

(i) In the latest series of models of particle acceleration in SNR the maximum rigidity is essentially higher than the 0.4 PV adopted by us from [1] and used as an argument for the knee position at ~3-4 PeV to be attributed to predominant oxygen ( Z = 8 ). Specifically, in [8] $R_{max}$= 15 PV, in [11] $R_{max}$= 40 PV, in [12] $R_{max}$= 10-100 PV for SNII expanding into the wind of the progenitor star, in [13] $R_{max}$= 90PV and in [14] $R_{max}$= 4·10$^6$ PV. One can see that the latest models of acceleration in SNR give the possibility of attributing the position of the knee to a lighter element than oxygen.

(ii) In [3] we analysed the possibility of different models of CR interaction giving a consistent result for the primary mass composition, measured by the mean logarithm of the mass, $\langle \ln A \rangle$, from the ratio of EAS muon and electron numbers $N_\mu / N_e$ and from the distribution of the depth of EAS maximum $X_{max}$. We concluded that the best consistency is for the QGSJET model. This conclusion has been confirmed by an analysis of other EAS characteristics [15]. Though this model is still imperfect [16], and needs further improvements, it is still the best. For this model the dominant element at the knee derived from the $N_\mu - N_e$ distribution obtained by KASCADE is He ( although there is ambiguity from the unfolding of the data into mass groups, see §1 ).

(iii) The $N_\mu - N_e$ - distribution in EAS observed at the mountain level has been analysed also by EAS-TOP group. Their final conclusion based both on EAS-TOP and MACRO experimental results is that the dominant component in the knee comprised helium nuclei [17,18]. However, the recent analysis of a similar $N_\mu - N_e$ distribution obtained by the GAMMA experiment at an even higher mountain altitude than EAS-TOP gives preference to dominant protons both for QGSJET and SIBYLL interaction models [19]. On the other hand, measurements at the highest altitude in Tibet in contrast with those just mentioned indicated that the knee is formed by nuclei heavier than helium [20]. Therefore the dispute on this problem is still going on.

(iv) In (10) we stressed the fact that according to the latest analysis of KASCADE data the mass composition changes very rapidly beyond the knee, becoming heavier. We have compared two versions of the Single Source model of the knee: with O or He dominant at the knee with these data. The result is shown in Figure 1. One can see that the dominance of He, its sharp cutoff and transition to dominant O in the region of the second peak (~12-16 PeV), gives a much faster rise of $\langle \ln A \rangle$ and fits the experimental data better.

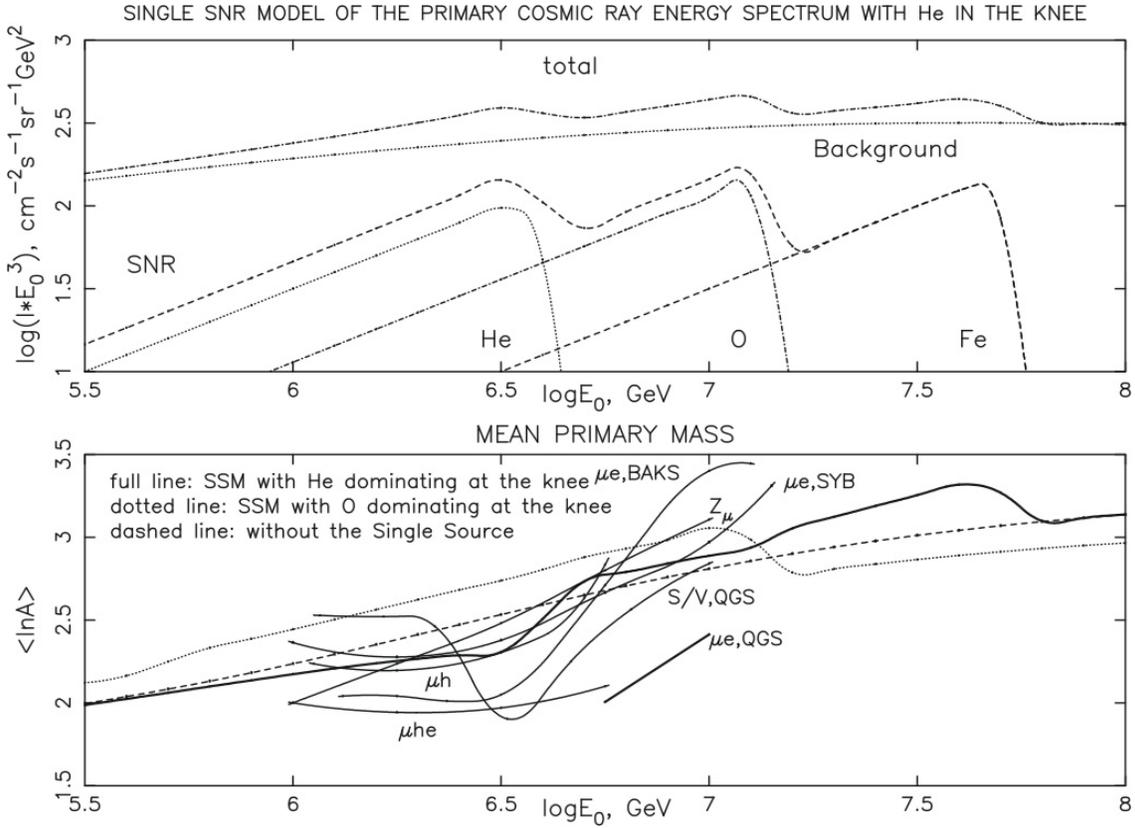

Figure 1. (a) The possible modification of the Single Source model of the knee assuming that helium and oxygen predominate in the first and second peaks of the intensity. The absolute intensity of the iron in the third peak is not determined from experiment unlike the intensities in the first and second peaks. (b) Mean logarithm of the mass number $\langle \ln A \rangle$ as a function of primary energy for the version of the Single Source model shown in the upper panel. Curves µeQGS and µeSYB show $\langle \ln A \rangle$ derived from EAS muons and electrons using the QGSJET and SYBILL interaction models, respectively [7], µh and µhe – from muons, hadrons and muons, hadrons and electrons, respectively [21], $Z_\mu$ from the longitudinal distribution of muon production heights [22] in the KASCADE experiment, µe,BAKS – from muons and electrons in the Baksan experiment [23], S/V,QGS –from the mean depth of shower maximum in SPASE-2/VULCAN experiment [24]. The rapid growth of $\langle \ln A \rangle$ is seen in all these experiments. Thick full and thin dotted lines show the energy dependence of the mean logarithm for two versions of the Single source model with helium and oxygen nuclei, predominant at the knee respectively.

(v) The analysis of the hadron energy $E_h$ in EAS detected by the ionization calorimeter at mountain level [25,26] showed the existence of remarkable peaks and troughs in the mean ratio $\langle E_h / N_e \rangle$. Their position ($N_e$), separation and the excursion from smooth dependence ( positive or negative ) analysed on the basis of simulations [27], which give $N_e(E_0,A)$ as a function of the energy $E_0$ and the mass A of the primary CR nucleus for the Tien-Shan observation level ( 3340 m a.s.l.), show better agreement with the predominance of He in the knee ( Figure 2 ).

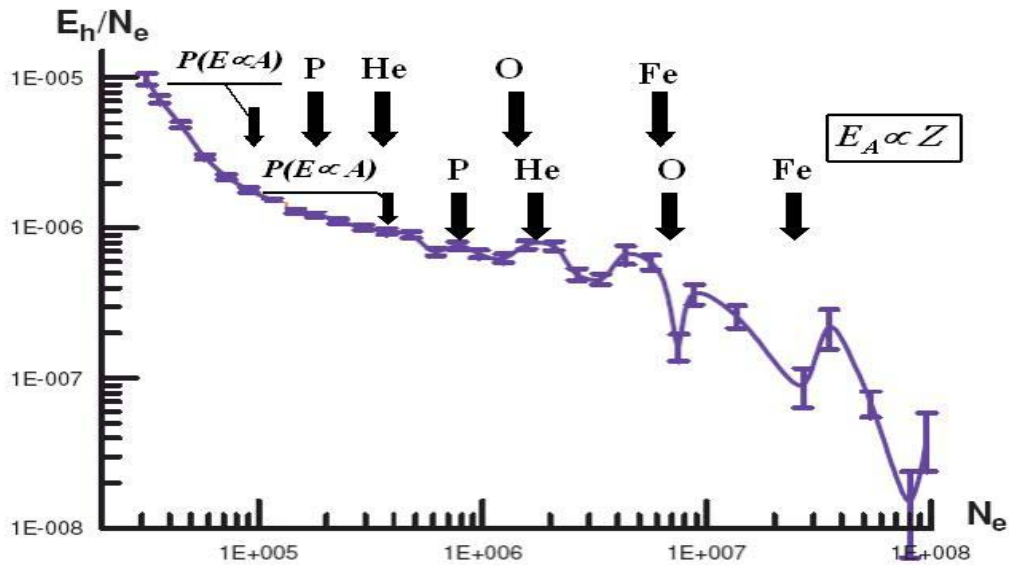

Figure 2. The energy of the EAS hadron component $E_h$, referred to the EAS size $N_e$, as a function of $N_e$ for EAS observed at mountain level [25,26]. The upper line of arrows indicates positions of energy cutoffs expected for the SSM with predominant oxygen (O) at the knee ( the knee at Tien-Shan level is observed at $\log N_e \approx 6.15$ ). The lower line of arrows indicates these positions for helium (He) predominant at the knee. The positions of the cutoffs were calculated on the basis of simulations [27] assuming that the values of energy cutoffs $E_A$ for different primary nuclei correspond to the same rigidity, ie. $E_A \propto Z$. The rigidity for the upper identifications is 0.4 PV and that for the lower, 1.5 PV. The position of the left arrow at both lines denoted as $P(E \propto A)$ corresponds to the position of the cutoff for protons expected if $E_A \propto A$.

(vi) Although it is too early to make a firm conclusion, due to unsufficient statistics, in the analysis of the mass composition made by KASCADE there is a hint for the existence of a 'third' peak of intensity [7]. The points at 50-70 PeV have a positive excursion from the smooth behaviour both for the QGSJET and SYBILL models, due presumably to Fe. The errors are still large but one has to keep this irregularity in mind and wait for the data from KASCADE-GRANDE.

There is another indication of the existence of the intensity peak at 50-70 PeV from the measurements of Cherenkov light in EAS made by the Tunka EAS array [28]. If the existence of this 'third' peak at 50-70 PeV is confirmed, then, taking into account the possibility of a small overestimation of the primary energy for cascades induced by the heavy nuclei such as Fe [29], it can provide strong evidence that the first peak at 3-4 PeV, ie. the knee, is due to the dominance of He.

vii) More important is the publication of new results from a direct measurement by 'ATIC' [6], confirmed at the latest 29$^{th}$ International Cosmic Ray Conference in Pune [30]. The data go up to $\sim \log E_n$ = 5.0 for protons, 4.3 for He-nuclei ($E_n$ being the energy per nucleon) and 3.5 for CNO and Fe. Interestingly, there is some evidence for an upturn in the helium spectrum starting at about $\log E_n$ = 4.0 such that **if** it is permitted to extrapolate by a factor 30 to reach the knee energy the necessary intensity is achieved. For CNO nuclei, where we drew attention to the flattening of the spectrum starting from $\log E_n$=2 [3] from the JACEE data, there is a smaller indication in the ATIC data. The situation is thus that the emphasis from direct measurements has moved somewhat from a CNO-knee to a He-knee; it is however, premature to be certain.

All these arguments taken separately are not overwhelming - just indications - but the total number of them is already large and growing with time, so that one has to take this 'collection of indications' seriously.

## 3. THE MONOGEM RING SNR

In a previous paper [4] we made the case for the Monogem Ring SNR being 'at the right distance and being of the right age' to satisfy our single source model. The calculations [4] showed agreement with the measured intensity, expressed in terms of rigidity, at the knee, assuming CNO for the component which gives the predominant contribution at the knee. However, since then we have found that the rigidity spectrum of CR expected from our Single Source had been overestimated. The true rigidity spectrum is lower in intensity and agreement between the model with oxygen predominant at the knee is damaged. It can be restored only for larger distances of the Single Source ( unless the propagation characteristics are changed ) or for smaller efficiency of the kinetic energy transfer to the accelerated SNR.

If, however, the particles are predominantly helium, rather than oxygen, the rigidity spectrum derived from the energy spectra of components in our Single Source will be a factor 4 more intense. However, this factor is not enough to compensate the excess of the expected intensity from the Monogem Ring over that derived from the Single Source model if the Monogem Ring had been created by a standard SN with an explosion energy of $10^{51}$erg, as had been assumed by us.

Interestingly, the quoted energy released by the Monogem Ring is lower: $1.9 \times 10^{50}$ erg, [31] than that of the standard $10^{51}$ erg. If this factor is taken into account, the agreement between the rigidity spectrum expected for the Monogem Ring SNR, with a distance of 290 pc and an age of 90 kyear, and the Single Source model, assuming helium predominant in the knee, is restored. The comparison between the rigidity spectra of particles expected from the Monogem Ring SNR and derived from the Single Source model, assuming oxygen or helium predominant at the knee, is shown in Figure 3. One can see that the agreement for the latter case is essentially better.

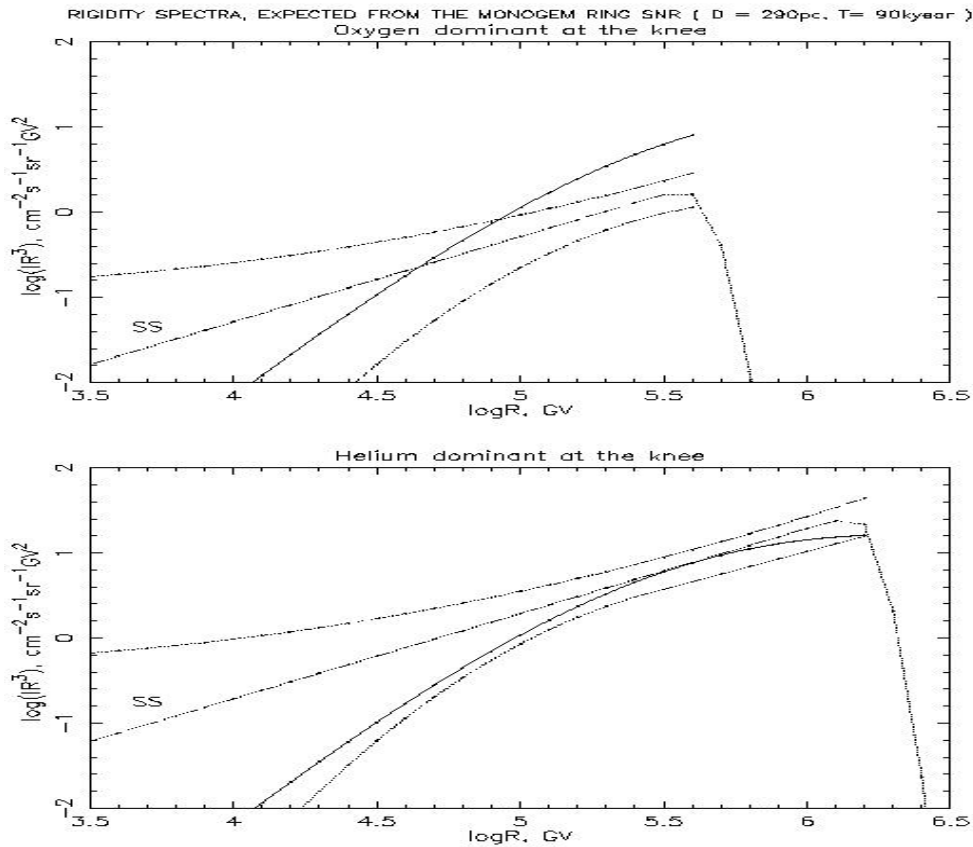

Figure 3. Comparison of the rigidity spectrum of particles expected from the Monogem Ring SNR with a distance of 290 pc, age – of 90 kyear, an explosion energy of $1.9 \times 10^{50}$ erg and that derived from the Single Source model assuming oxygen ( upper panel ) or helium ( lower panel ) predominant at the knee. There is clearly a consistent picture for the knee being due, principally, to helium nuclei being accelerated by the Monogem Ring SNR at the currently predicted distance and age [5].

In principle, it should also be possible to show that a specific model gives the correct rigidity ( or energy ) cut-off. i.e. the maximum rigidity attainable [8,11,12,13,14,32], since all of them consider SNR as the energy source for CR. For the more conservative models with the SN shock accelerating CR from an ordinary ISM [33,34,35,36] there is at first sight an inconsistency between the values of the CR intensity predicted and the maximum rigidity: SNR such as the Monogem Ring cannot give such a high maximum rigidity as 1.5-1.6 PV for such a low explosion energy as $0.19 \times 10^{51}$ erg. However, the answer seems to lie with the clumpiness of the remnant. The models invariably assume spherical symmetry, in contrast to the actual situation. In the dense knots of actual SNR [5,29] it is presumably the volume density of the shock energy, etc. that is important rather than the spherically averaged value, with the result that there is decoupling between the maximum energy ( governed by the local situation in the knot ) and the total intensity of CR produced ( governed by the total energy of the SNR, the fraction of energy going into CR and the spectral shape ).

4. **CONCLUSION**

Concerning our identification of the knee as indicating a significant contribution from a local, single, recent SN, our view is unchanged. Turning to the less firm estimate of the mass of particles from the single source at the knee, new evidence gives rather more credence to the suggestion that helium rather than oxygen nuclei are responsible for the knee. The evidence comes from hints from new direct spectral measurements and several indirect indications, both experimental and theoretical. The evidence is not proof, but the growing number of indications allows one to consider this suggestion as a likely option.